\documentclass[%
 reprint,
 superscriptaddress,
nobibnotes,
 amsmath,amssymb,
 aip,
]{revtex4-2}

\usepackage{graphicx}
\usepackage{dcolumn}
\usepackage{bm}

\usepackage{physics}
\usepackage{hyperref}
\usepackage{float}
\usepackage{xcolor}

\begin{document}


\title{Deep Learning Option Pricing with Market Implied Volatility Surfaces}

\author{Lijie Ding}
\email{dingl1@ornl.gov}
\affiliation{Neutron Scattering Division, Oak Ridge National Laboratory, Oak Ridge, TN 37831, USA}
\author{Egang Lu}
\affiliation{Independent researcher}
\author{Kin Cheung}
\affiliation{Valkin Holdings, LLC, Rowland Heights, CA 91748, USA}
\date{\today}

\begin{abstract}
We present a deep learning framework for pricing options based on market-implied volatility surfaces. Using end-of-day S\&P 500 index options quotes from 2018-2023, we construct arbitrage-free volatility surfaces and generate training data for American puts and arithmetic Asian options using QuantLib. To address the high dimensionality of volatility surfaces, we employ a variational autoencoder (VAE) that compresses volatility surfaces across maturities and strikes into a 10-dimensional latent representation. We feed these latent variables, combined with option-specific inputs such as strike and maturity, into a multilayer perceptron to predict option prices. Our model is trained in stages: first to train the VAE for volatility surface compression and reconstruction, then options pricing mapping, and finally fine-tune the entire network end-to-end. The trained pricer achieves high accuracy across American and Asian options, with prediction errors concentrated primarily near long maturities and at-the-money strikes, where absolute bid-ask price differences are known to be large. Our method offers an efficient and scalable approach requiring only a single neural network forward pass and naturally improve with additional data. By bridging volatility surface modeling and option pricing in a unified framework, it provides a fast and flexible alternative to traditional numerical approaches for exotic options.
\end{abstract}
\maketitle


\section{Introduction}
Options are fundamental financial derivatives that enable investors to hedge risks, speculate on market movements, and manage portfolios in increasingly complex markets~\cite{BlackScholes1973}. The global options market has grown significantly, driven by demand for sophisticated instruments like American puts and arithmetic Asian options\cite{wilmott2013paul,shreve2004stochastic}. However, pricing option portfolios often poses challenges. Traditional parametric models, such as the Black-Scholes framework, assume constant volatility and log-normal dynamics. Such assumptions necessitate model adjustments in the real-world with volatility smiles, term structure, and jumps~\cite{Hull2017,cont2003financial}. Numerical methods, such as Monte Carlo simulations or finite difference schemes, offer flexibility for exotics but are computationally intensive, particularly when handling high-dimensional market-implied volatility surfaces calibrated from liquid listed European options~\cite{Glasserman2004,duffie2010dynamic}. As the complexity of these surfaces and the variety of exotic instruments grow, traditional approaches struggle to scale, limiting their efficiency for real-time pricing and risk assessment.

Recent advances in machine learning (ML)\cite{murphy2012machine,goodfellow2016deep} have revolutionized financial modeling by offering data-driven, non-parametric alternatives that adapt to complex market patterns directly \cite{Dixon2020, Liu2019, Hirsa2019}. For options pricing, Physics-informed neural network\cite{raissi2019physics} has been applied on solving the partial differential equations by including the Black-Scholes and related equations into the loss function\cite{gatta2023meshless, hainaut2024option,wang2023deep,bai2022application}. Although this approach requires no pricing data for training, it is limited by a pre-fixed set of  parameters for each training. Other approaches use pricing data to train ML models such as neural network or Gaussian process regressor to learn the direct mapping from the pricing inputs (e.g. strike, time to maturity, volatility) to the pricing results\cite{de2018machine,ndikum2020machine,gaspar2020neural,anderson2023accelerated}. However, these approaches have typically used flat volatility without accounting for the complexities of volatility surfaces observed from actual market data\cite{cao2021deep,ruf2019neural,bloch2019option,culkin2017machine}, limiting their practicality for real world applications. 

In this work, we introduce a variational autoencoder (VAE)\cite{doersch2016tutorial,pu2016variational,bergeron2021variational}-based neural network framework to price options directly from complete sets of market-implied volatility surfaces. Using daily historical end-of-day S\&P 500 European options data from 2018 to 2023, we construct arbitrage-free volatility surfaces on a $41\times 20$ grid of log-moneyness and time-to-maturity. The VAE compresses these high-dimensional surfaces into a 10-dimensional latent space, a multilayer perceptron (MLP) then maps these latent variables, along with strike and maturity, to prices for American puts and arithmetic Asian options\cite{vecer2001new}. This approach overcomes the limitations of parameterization models\cite{gatheral2014arbitrage, ding2025fast} by using a data-driven approach that bypasses the computationally intensive numerical solvers, thus offering an efficient, GPU-parallelization compatible pricing methodology that can scale further with increasingly available market data.

\section{Method}
\label{sec:method}

\subsection{Data preparation}
To prepare the data for training our deep learning model, we first create implied volatility surface data from the end-of-day European options pricing data collected from the freely available optionsDX platform. We then use these volatility surfaces as inputs to generate prices for American puts and arithmetic Asian options using the corresponding pricing engines from the publicly available QuantLib library\cite{varma2015derivatives,varma2016computational}.

For each business day in our data series, we generate the market implied volatility surface from a chain of options pricing data using the standard Black-Scholes pricing formula. For each options pricing data point, we compute the corresponding implied volatility $\sigma_{BS}$ by solving the $\sigma_{BS}$ from the pricing formula of European options:
\begin{equation}
    \begin{aligned}
        c(K,T) &= S(0)N(d_1) - e^{-rT}KN(d_2) \\
        p(K,T) &= e^{-rT}KN(-d_2) - S(0)N(-d_1) \\    
        d_1 &= \frac{\log{(S/K)}+(r+\sigma_{BS} ^2 /2)T}{\sigma_{BS} \sqrt{T}} \\
        d_2 &= d_1 - \sigma_{BS} \sqrt{T}
    \end{aligned}
\end{equation}
where $c(K,T)$ and $p(K,T)$ are the price of the call and put option, respectively, at strike $K$ and time to maturity $T$. $S(0)$ is the spot value of the underlying, $r$ is the risk-free rate, $N(\cdot)$ is the standard normal cumulative distribution function. For any strike $K$ and time to maturity $T$ where we have both put and call prices, we only use the implied volatility computed from the option that is out-of-the-money, i.e., we use the implied volatility of the put option for $K<S(0)$ and that of the call option for $K>S(0)$, respectively, as that is usually the option that is more liquid.

\begin{figure}
    \centering
    \includegraphics[width=\linewidth]{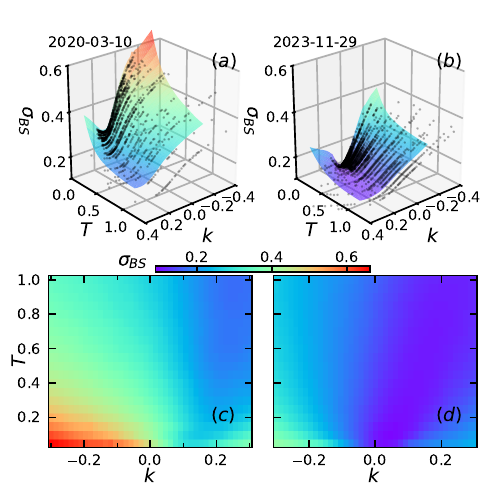}
    \caption{Illustration of the market implied volatility surfaces for SPX (Standard and Poor's 500 index). (a) Black-Scholes volatility $\sigma_{BS}$ versus log moneyness $k=\log(K/S_0)$ and time to maturity in year $T$ for Mar 10th, 2020, the day after the Covid crash. Gray dots are marker observed point, color surface is interpolated plane. (b) volatility surface at a random day. (c) and (d) are the heat map view of volatility surfaces in (a) and (b), respectively.}
    \label{fig:vol_illustration}
\end{figure}

To further standardize the format of the volatility surfaces for efficient machine training, we calculate the $\sigma_{BS}(k,T)$ surface on a grid fixed in log moneyness $k=\log(K/S_0)$ and time to maturity $T$ by interpolation using neighboring data points, resulting in a uniform $41\times 20$ matrix for each business day's worth of data, with 41 log moneyness $k \in [-0.3,0.3]$ and 20 option maturities in years $T\in[0.05, 1]$. The scatter points in Fig.~\ref{fig:vol_illustration}shows two such sample implied volatility surfaces and the corresponding heat maps as observed from the market for SPX. Due to the limited data quality for some of the end-of-day quotes, some interpolated volatility surfaces appear not to be arbitrage-free. As a result, we have to filter out and eliminate these surfaces after carrying out test pricing on vanilla option valuation using QuantLib. In total, we have collected 1051 arbitrage-free volatility surfaces $\vb{F} = \{\sigma_{BS}(k,T)\}$ between 2018 to 2023, and we subsequently divide $\vb{F}$ into a training set $\{\sigma_{BS}(k,T)\}_{train}$ containing 840 volatility surfaces and a testing set $\{\sigma_{BS}(k,T)\}_{test}$ containing 211 volatility surfaces.

To generate pricing data for American puts and arithmetic Asian (call and put) options, we use the QuantLib library to compute option prices based on the market-implied volatility surfaces as ground truth valuations. We generate random combinations of log moneyness $k$ and time to maturity $T$ from uniformly distributed domains of $k \in [-0.3,0.3]$ and $T\in[0.05, 1]$ along with randomly chosen quote dates. Then the entire volatility surface for the corresponding date is passed onto the QuantLib pricer along with the strike $K=S_0 e^k$ and time to maturity $T$. For American puts, we have generated 20,000 price data using the volatility surface $\sigma_{BS}(k,T)$ from the training set, and 4,000 pricing data using the $\sigma_{BS}(k,T)$ from the testing set. For the arithmetic Asian options, we have prepared 10,000 pricing data for both calls and puts as training set, and 2,000 pricing data for each of calls and puts as testing data.

\subsection{Variational Autoencoder-based neural network}
\label{ssec:vae}
\begin{figure}
    \centering
    \includegraphics[width=\linewidth]{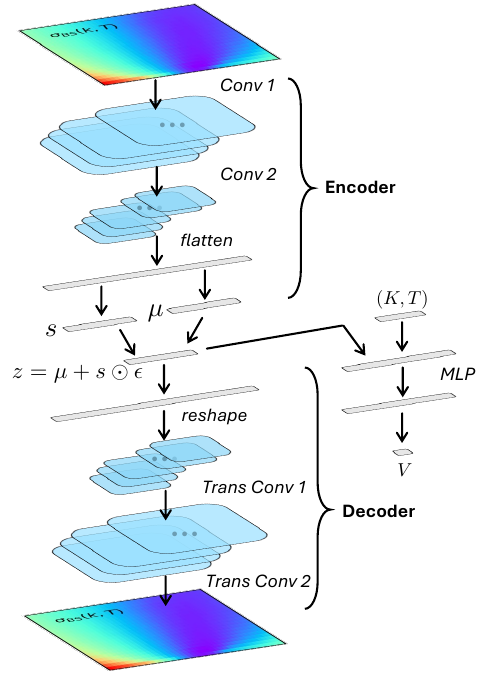}
    \caption{Architecture of the variational autoencoder (VAE)-based neural network pricer. The neural network consists of two main parts: a VAE that extract the latent variables of the volatility surfaces, and a multilayer perception (MLP) that maps the instrument variables and the latent variables of the volatility surfaces to the option price. The VAE is made of an encoder of 2 convolutional layers, a latent space with 10 dimensions, followed by a decoder of 2 transposed convolutional layers symmetric to the encoder.}
    \label{fig:NN_architecture}
\end{figure}

To learn the mapping between the option prices and the pricing inputs, which include the strike, the time to maturity, and the entire volatility surface (rather than a single interpolated volatility input), we designed a VAE-based neural network as illustrated in Fig.~\ref{fig:NN_architecture}. As shown in sec~\ref{sec:results}, this approach is based on the observation that even though the volatility surface representation $\sigma_{BS}(k,T)$ is high dimensional in nature, the overall shape of the $\sigma_{BS}(k,T)$ can be captured with a set of much lower dimensional latent variables. 

Based on this principle, the left part of the neural network is a VAE made of three main components: an encoder, a latent space, and a decoder. The encoder will compress each input volatility surface $\sigma_{BS}(k,T)$, a  $41\times20$ matrix, into a 10 dimensional latent space, and the decoder will reconstruct the volatility surface $\sigma'_{BS}(k,T)$ back from the latent space. The encoder consists of two convolutional layers, each with kernel size 3 and strides 2. The first convolutional layer has 16 channels, the second one has 32 channels. After passing the two layers, the $41\times20$ $\sigma_{BS}(k,T)$ will be transformed into a $11 \times 5\times32=1760$ dimensional vector, which is then mapped onto the 10 latent variables, each with mean $\mu_{i}$'s and standard deviation $s_{i}$'s. These variables are rewritten as $z_{i} = \mu_{i} + s_{i}\odot \epsilon$ with normally distributed variable $\epsilon\sim \mathcal{N}(0,1)$. Values for the 10 latent variables $z$  are obtained after randomly sampled from $\epsilon$, they will then go through the decoder which generates a reconstructed volatility surface $\sigma'_{BS}(k,T;\epsilon)$ as output, and this procedure is repeated 10 times, each time with values of $z$'s generated from another randomly sampled $\epsilon$. The volatility surface averaged over these 10 outputs $\sigma'_{BS}(k,T) = \left< \sigma'_{BS}(k,T;\epsilon)\right>_\epsilon$ is the reconstructed volatility surface used in the loss function. The VAE is then trained to minimize the following loss function that calculates the mean squared error between the input and output volatility surfaces, with $N$ being the number of volatility surfaces:
\begin{equation}
    L_{VAE} =\frac{1}{N} \sum_{\sigma_{BS}(k,T)} \left<\left[\sigma_{BS}(k,T) - \sigma'_{BS}(k,T) \right]^2 \right>_{k,T}
    \label{equ:loss_VAE}
\end{equation}

To learn the price of the options for each set of inputs $(\sigma_{BS}(k,T),K,T)$, the volatility surface is fed into the encoder so that a set of latent variables is generated. These latent variables are then entered into the MLP on the right side of the neural network as in Fig.~\ref{fig:NN_architecture}, along with the instrument parameters such as strike $K$ and time to maturity $T$, and an option price $V'$ is generated. The pricer MLP is then trained to minimize the following loss function:
\begin{equation}
    L_{MLP} =\frac{1}{M} \sum_{V}(V' - V)^2
    \label{equ:loss_MLP}
\end{equation}
where $M$ is the number of pricing data corresponding to each combination of $(\sigma_{BS}(k,T),K,T)$ and V is the ground truth option price as computed with the same inputs into QuantLib.

To train the entire neural network, we need to first train the VAE until it is able to reconstruct volatility surfaces satisfactorily. During this first stage of training, the MLP is not involved. In the second stage, We focus on training the MLP. During this process, we use the trained encoder to generate the latent variables to be fed into MLP, but the encoder itself is not part of the training, only the MLP is being trained. Lastly, in the fine tuning process, we train both the encoder and the MLP together as driven by MLP loss Eq.~\eqref{equ:loss_MLP}. In practice, the neural network is implemented using PyTorch and trained using Adam optimizer with CosineAnnealingLR scheduler. We train the VAE for 3,000 epoch, the MLP for 150 epoch and fine tune for another 50 epoch.

\section{Results}
\label{sec:results}

We first demonstrate the feasibility of the dimension reduction of the market implied volatility surface data. We then discuss the training of our deep learning model and carry out analysis of the trained model. Finally, we apply our trained model as a pricer for American puts and arithmetic Asian options, both of which do not have closed form solutions.

\subsection{Feasibility of dimension reduction of volatility surfaces}
We first inspect the dataset $\vb{F} = \{\sigma_{BS}(k,T)\}$ of all 1,051 SPX volatility surfaces we collected. By rearranging each volatility surface $\sigma_{BS}(k,T)$'s  $41\times 20$ matrix components into a column vector, $\vb{F}$ becomes a $1051\times 820$ matrix with each column representing the volatility surface for a given date. We then carry out principal component analysis of $\vb{F}$ using singular value decomposition such that $\vb{F} = \vb{U}\Sigma \vb{V}^T$, in which the $\vb{U}$ is $1051\times 1051$, $\Sigma$ is $1051\times 820$, and $\vb{V}^T$ is $820\times 820$. The diagonal entries of the $\Sigma$ are the singular values that determine the projection of $\vb{F}$ onto the singular vector space $\vb{V}$.

\begin{figure}[!h]
    \centering
    \includegraphics[width=\linewidth]{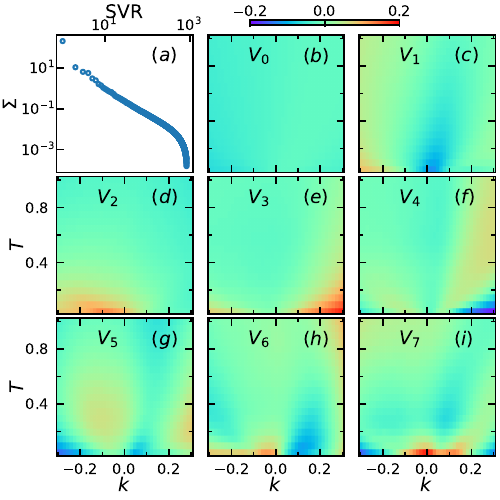}
    \caption{Principal component analysis of the collected SPX volatility surfaces using singular value decomposition (SVD). (a) Decay of the singular value entry in $\Sigma$ with $\vb{F} = \vb{U}\Sigma \vb{V}^T$ versus the singular value rank (SVR). (b)-(i) The first 8 singular vectors of the volatility dataset $\vb{F}$.}
    \label{fig:vol_surface_svd_analysis}
\end{figure}

As shown in the log-log scale plot in Fig.~\ref{fig:vol_surface_svd_analysis}(a), the singular values decay rapidly as the rank increases, meaning that the higher rank singular vectors in $\vb{V}$ becomes significantly less important in the representation of the volatility surfaces data $\vb{F}$, thus it is possible to perform dimensional reduction for the $41\times 20$ shape volatility surface data\cite{ackerer2020deep}. Fig.~\ref{fig:vol_surface_svd_analysis}(b)-(i) shows the first 8 singular vectors in $\vb{V}$, reshaped back to the $41\times 20$ representation for easier comparison with the volatility surface data. The first and most dominant singular vector Fig.~\ref{fig:vol_surface_svd_analysis}(b) is relatively uniform and flat across strikes and time to maturity, meaning it represents the overall parallel shifts of the entire volatility surface. The blue shade in Fig.~\ref{fig:vol_surface_svd_analysis}(c) shows how volatility tends to decrease for higher strike options with time to maturity, and the brown shade shows the increases for the lower strikes options. Together they can be interpreted as the opening and the closing of the volatility smile as a function of maturity. Fig.~\ref{fig:vol_surface_svd_analysis}(d) and (e) mostly adjust  the skew at the wings (high strike calls and low strike puts) of the volatility surface. And as the singular value rank increases, the pattern becomes more detailed, meaning each higher rank singular vector is representing ever more subtle aspects of the volatility surface, and are increasing less important. The rapid decay suggests 5--10 dimensions may suffice, motivating our VAE choice.

\subsection{Neural network training and analysis}
\begin{figure}[!h]
    \centering
    \includegraphics[width=\linewidth]{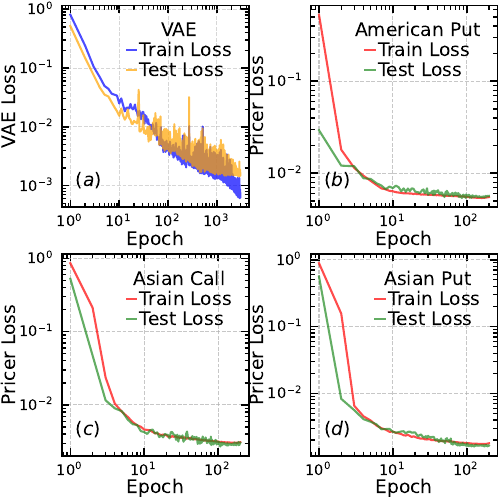}
    \caption{Loss curves for the training of the neural-network model. (a) Train and test loss for training the variational autoencoder using only the volatility surfaces data. (b)-(d) Train and test loss for training the pricing network using the pricing data consist of both the volatility surface and instrument parameters.}
    \label{fig:loss_curves}
\end{figure}

In order to train the neural network model to price options using the entire volatility surface as input, we first need to achieve the dimension reduction of the volatility surface. As a result, we must first train the VAE on the left side of Fig.~\ref{fig:NN_architecture} until sufficiently accurate latent variables can be extracted from the encoder. Fig.~\ref{fig:loss_curves}(a) shows the loss curve of the VAE loss $L_{VAE}$ versus training epoch. Training loss decreases steadily, while testing loss plateaus in the early thousands. We stopped at 3,000 epoch as the testing loss curve indicates that the network is well-trained without overfitting at that point. After the VAE is trained, we use the encoder to generate the latent variables that are fed into MLP along with K and T to train the option pricer. Fig.~\ref{fig:loss_curves}(b)-(d) show the loss curve for American put, Asian call, and Asian put, respectively. Again we stopped the training when the test loss curve seems to have reached a plateau. These loss curves are all well-behaved and indicate that our training steps are efficient and sufficient.

\begin{figure}
    \centering
    \includegraphics[width=\linewidth]{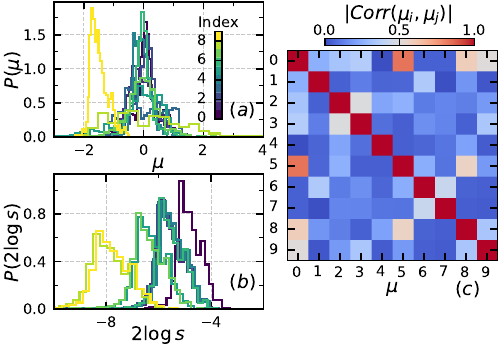}
    \caption{Distribution and statistics of latent variable of the training set of volatility surface. (a) Density distribution $P(\mu)$ of all 10 mean latent variables. (b) Density distribution of variance latent variables. (c) Absolute value of the correlation between all mean latent variables $(\mu_i,\mu_j)$.}
    \label{fig:latent_variables}
\end{figure}

To further examine the trained neural network, we look at the distribution of the latent variables from all of the training volatility surfaces. Fig.~\ref{fig:latent_variables}(a) and (b) show the distribution of the mean and log variance latent variables. Most of the mean latent variables $\mu$ are distributed around 0 with one around $-1$, while the log variance latent variable $2\log s$ are mostly very small, at least smaller than $-4$, which results in nearly $s\simeq 0$. The main takeaway from this comparison is that most of the information is stored in the mean latent variables $\mu$'s as their magnitudes dominates those of the variances. In addition, when looking at the correlation of the 10 $\mu$'s in Fig.~\ref{fig:latent_variables}(c), where the absolute value of the correlations between different indices of $\mu$ is shown, the heat map shows emerging correlation between different indices of $\mu$, such as between $\mu_{0}$ and $\mu_{5}$, indicating the number of latent space dimension is starting to saturate, and that our choice of having 10 variables appear sufficient for our dataset.

\begin{figure}
    \centering
    \includegraphics[width=\linewidth]{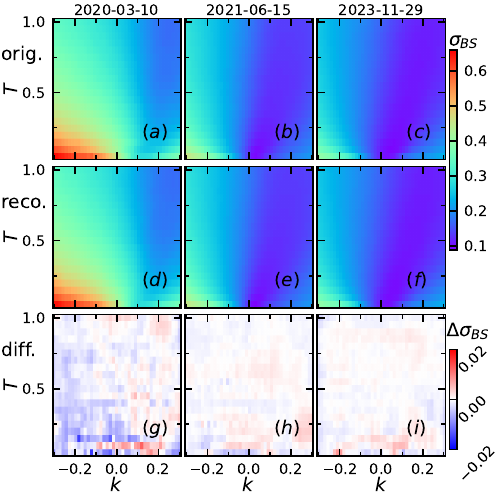}
    \caption{Sample reconstruction of the volatility surface using the variational autoencoder network. (a)-(c) Original volatility surface of 3 different quote dates from the volatility dataset. (d)-(f) Reconstruction of the volatility surface for (a)-(c), respectively. (g)-(i) Difference $\delta \sigma_{BS} =  \sigma'_{BS} - \sigma_{BS}$ between the reconstruction $\sigma'_{BS}$ and original volatility surface $\sigma_{BS}$.}
    \label{fig:vae_vol_reconstructions}
\end{figure}

Furthermore, Fig.~\ref{fig:vae_vol_reconstructions} demonstrates the reconstruction of the volatility surfaces using the trained VAE. The first row shows the original volatility surfaces $\sigma_{BS}(k,T)$ for three different quote dates, including the day after the Covid crash ( Fig.~\ref{fig:vae_vol_reconstructions}(a)). The second row shows the VAE-reconstructed volatility surfaces $\sigma'_{BS}(k,T)$, and the third row shows the difference $\Delta\sigma_{BS} = \sigma'_{BS}(k,T) - \sigma_{BS}(k,T)$. The differences $\Delta\sigma_{BS}$ are relatively small in Fig.~\ref{fig:vae_vol_reconstructions}(h) and (i) as these two dates are fairly typical, while the differences $\Delta\sigma_{BS}$ are slightly larger for Fig.~\ref{fig:vae_vol_reconstructions}(g). However, given the extreme volatile environment resulting from the Covid driven crash, the reconstructed volatility surface is still quite accurate.

\subsection{Evaluation of neural network pricer}
Finally, we apply the trained neural network pricer to price American Puts and arithmetic Asian options, both lacking closed-form solutions and must be computed numerically. Both the training and testing data sets are generated from historical volatility surfaces, and ground truth valuations are obtained from Quantlib.  The training and testing sets are independent, i.e., the neural network was not trained with any volatility surfaces from the set of testing pricing data. 

\begin{figure}[!h]
    \centering
    \includegraphics[width=\linewidth]{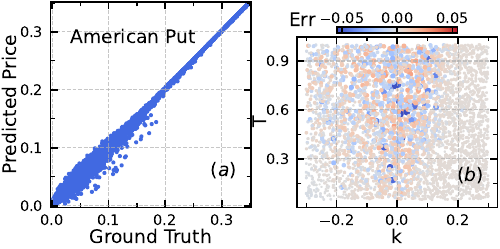}
    \caption{Benchmark the performance of the neural network pricer for American Put. (a) Comparison between the ground truth price $P$ and neural network-predicted price $P'$. (b) Distribution of $Err = P'-P$ of all test pricing data in the log moneyness $k$ and time to maturity $T$ plane.}
    \label{fig:AmericanPut_price_prediction}
\end{figure}

As shown in Fig.~\ref{fig:AmericanPut_price_prediction}(a) for American Put prices, the neural network predicted prices align well with the ground truth prices, with all of the predicted prices distributing nicely around the diagonal line of perfect matches. Fig.~\ref{fig:AmericanPut_price_prediction}(b) shows a detailed breakdown of the distribution of prediction errors $Err=V'-V$ in the log moneyness $k$ and time to maturity $T$ plane. The errors are overall small, with the  few larger errors observed along the longer expiries and close to at-the-money strikes, where absolute prices are low. 

\begin{figure}[!h]
    \centering
    \includegraphics[width=\linewidth]{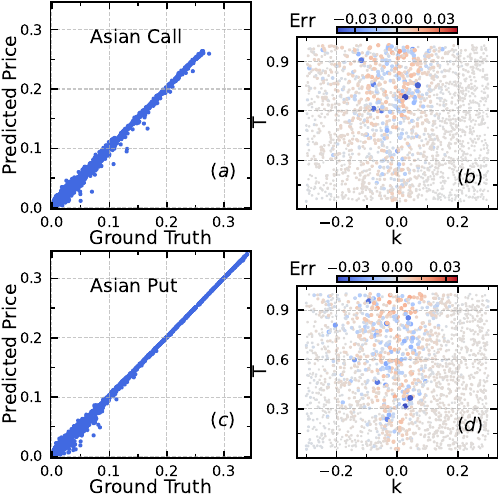}
    \caption{Benchmark the performance of the neural network pricer for Asian call and Asian put. (a) Comparison between the ground truth price $V$ and neural network-predicted price $V'$ for Asian call. (b) Distribution of $Err = V'-V$ of all test pricing data in the log moneyness $k$ and time to maturity $T$ plane for Asian call. (c)-(d) Similar to (a) and (b) but for Asian put.}
    \label{fig:AsianOpt_price_prediction}
\end{figure}

Moreover, Fig.~\ref{fig:AsianOpt_price_prediction}(a) and (c) show the comparison between the neural network predicted prices versus the ground truth prices for arithmetic Asian call and put options. For both types of Asian options, the prices again agree very well. Similarly, Fig.~\ref{fig:AsianOpt_price_prediction}(b) and (d) show the error in the $k$ and $T$ plane. Similar to the American Puts, with the exception of a few outliers close to at-the-money strikes where the absolute prices are low, most of numerical errors are quite modest.

\section{Summary}
In this work, we introduce a deep learning approach for options pricing that incorporates the entire volatility surface and bypasses traditional numerical methods. Our model is trained on market-implied volatility surfaces from SPX data, with pricing data generated using the QuantLib package. During the pricing, the entire volatility surface is used instead of using a single interpolated volatility input. We demonstrate that the two-dimensional volatility surface can be compressed into a lower-dimensional representation via singular value decomposition. Building upon this insight, we employ a VAE-based neural network to reduce each volatility surface specified by $820$-dimensions into a $10$-dimensional latent space represented by the neural network and the 10 sets of latent variables. After we have confirmed that the VAE is able to successfully and accurately reconstruct full volatility surfaces from the latent space, we use the encoder portion of the VAE to generate the latent variables, then together with option parameters such as strike and maturity, they are fed into an MLP to compute option prices. We have applied this methodology to American puts and arithmetic Asian options, whose valuations typically require lengthy numerical calculations. The results show that our neural network achieves high accuracy across all three types of options.

Our market data driven approach is efficient, scalable, and flexible. Unlike numerical methods, it computes exotic prices in a single forward pass and supports GPU-parallel processing of many trades. Demonstrated on end-of-day SPX data with fixed model size and latent dimensions, the architecture scales easily to broader resolutions by adjusting model size. Our deep learning framework will also improve as more data becomes available, making it increasingly accurate and adaptable over time. More importantly, this methodology can be used to compute portfolios of exotic options extremely quickly, thus providing traders with live risk management and intra-day monitoring of profit and loss capabilities. Instead of using QuantLib, the architecture can easily accommodate proprietary pricing models to generate ground truth valuations during training such that the trained neural network can generate fast options valuations that are consistent with one's proprietary models, no matter how slow and complicated the proprietary models may be. 

Looking ahead, our framework can be extended in several directions. One natural step is to broaden the dataset to include additional equity indices and single-stock options, as well as broaden the types of exotic options and structured products to be evaluated. Beyond equities, the same methodology can also be adapted to other asset classes such as commodities, foreign exchange, cryptocurrency and some interest rate derivatives where options pricing also relies similarly on using such volatility surfaces as inputs. Finally, the VAE-based approach itself offers a promising path toward building a flexible, accurate and yet tractable volatility surface representation with substantially reduced dimensions\cite{ning2023arbitrage,wang2025controllable,kelly2023deep}. It will be interesting to see if this approach can be applied to more complicated volatility inputs such as strike dependent swaption volatility surfaces\cite{fan2003hedging} that are used with fixed income derivatives models such as SABR\cite{hagan2015probability}. 

\section*{Data Availability}
The code and data for this work are available at the GitHub repository \url{https://github.com/ljding94/VAE_pricing}

\section*{Author Contributions}
LD led the research. LD and EL prepared the market data. LD derived the theoretical framework, developed the code, generated and analyzed the data; and LD, EL, and KC wrote and edited the manuscript.

\section*{Acknowledgment}
This research was sponsored by the Laboratory Directed Research and Development Program of Oak Ridge National Laboratory, managed by UT-Battelle, LLC, for the U.S. Department of Energy.


\section*{References}
\bibliography{reference}

\begin{thebibliography}{39}%
\makeatletter
\providecommand \@ifxundefined [1]{%
 \@ifx{#1\undefined}
}%
\providecommand \@ifnum [1]{%
 \ifnum #1\expandafter \@firstoftwo
 \else \expandafter \@secondoftwo
 \fi
}%
\providecommand \@ifx [1]{%
 \ifx #1\expandafter \@firstoftwo
 \else \expandafter \@secondoftwo
 \fi
}%
\providecommand \natexlab [1]{#1}%
\providecommand \enquote  [1]{``#1''}%
\providecommand \bibnamefont  [1]{#1}%
\providecommand \bibfnamefont [1]{#1}%
\providecommand \citenamefont [1]{#1}%
\providecommand \href@noop [0]{\@secondoftwo}%
\providecommand \href [0]{\begingroup \@sanitize@url \@href}%
\providecommand \@href[1]{\@@startlink{#1}\@@href}%
\providecommand \@@href[1]{\endgroup#1\@@endlink}%
\providecommand \@sanitize@url [0]{\catcode `\\12\catcode `\$12\catcode
  `\&12\catcode `\#12\catcode `\^12\catcode `\_12\catcode `\%12\relax}%
\providecommand \@@startlink[1]{}%
\providecommand \@@endlink[0]{}%
\providecommand \url  [0]{\begingroup\@sanitize@url \@url }%
\providecommand \@url [1]{\endgroup\@href {#1}{\urlprefix }}%
\providecommand \urlprefix  [0]{URL }%
\providecommand \Eprint [0]{\href }%
\providecommand \doibase [0]{https://doi.org/}%
\providecommand \selectlanguage [0]{\@gobble}%
\providecommand \bibinfo  [0]{\@secondoftwo}%
\providecommand \bibfield  [0]{\@secondoftwo}%
\providecommand \translation [1]{[#1]}%
\providecommand \BibitemOpen [0]{}%
\providecommand \bibitemStop [0]{}%
\providecommand \bibitemNoStop [0]{.\EOS\space}%
\providecommand \EOS [0]{\spacefactor3000\relax}%
\providecommand \BibitemShut  [1]{\csname bibitem#1\endcsname}%
\let\auto@bib@innerbib\@empty
\bibitem [{\citenamefont {Black}\ and\ \citenamefont
  {Scholes}(1973)}]{BlackScholes1973}%
  \BibitemOpen
  \bibfield  {author} {\bibinfo {author} {\bibfnamefont {F.}~\bibnamefont
  {Black}}\ and\ \bibinfo {author} {\bibfnamefont {M.}~\bibnamefont
  {Scholes}},\ }\bibfield  {title} {\enquote {\bibinfo {title} {The pricing of
  options and corporate liabilities},}\ }\href@noop {} {\bibfield  {journal}
  {\bibinfo  {journal} {Journal of Political Economy}\ }\textbf {\bibinfo
  {volume} {81}},\ \bibinfo {pages} {637--654} (\bibinfo {year}
  {1973})}\BibitemShut {NoStop}%
\bibitem [{\citenamefont {Wilmott}(2013)}]{wilmott2013paul}%
  \BibitemOpen
  \bibfield  {author} {\bibinfo {author} {\bibfnamefont {P.}~\bibnamefont
  {Wilmott}},\ }\href@noop {} {\emph {\bibinfo {title} {Paul Wilmott on
  quantitative finance}}}\ (\bibinfo  {publisher} {John Wiley \& Sons},\
  \bibinfo {year} {2013})\BibitemShut {NoStop}%
\bibitem [{\citenamefont {Shreve}\ \emph {et~al.}(2004)\citenamefont {Shreve}
  \emph {et~al.}}]{shreve2004stochastic}%
  \BibitemOpen
  \bibfield  {author} {\bibinfo {author} {\bibfnamefont {S.~E.}\ \bibnamefont
  {Shreve}} \emph {et~al.},\ }\href@noop {} {\emph {\bibinfo {title}
  {Stochastic calculus for finance II: Continuous-time models}}},\
  Vol.~\bibinfo {volume} {11}\ (\bibinfo  {publisher} {Springer},\ \bibinfo
  {year} {2004})\BibitemShut {NoStop}%
\bibitem [{\citenamefont {Hull}(2017)}]{Hull2017}%
  \BibitemOpen
  \bibfield  {author} {\bibinfo {author} {\bibfnamefont {J.~C.}\ \bibnamefont
  {Hull}},\ }\href@noop {} {\emph {\bibinfo {title} {Options, Futures, and
  Other Derivatives}}},\ \bibinfo {edition} {9th}\ ed.\ (\bibinfo  {publisher}
  {Pearson},\ \bibinfo {year} {2017})\BibitemShut {NoStop}%
\bibitem [{\citenamefont {Cont}\ and\ \citenamefont
  {Tankov}(2003)}]{cont2003financial}%
  \BibitemOpen
  \bibfield  {author} {\bibinfo {author} {\bibfnamefont {R.}~\bibnamefont
  {Cont}}\ and\ \bibinfo {author} {\bibfnamefont {P.}~\bibnamefont {Tankov}},\
  }\href@noop {} {\emph {\bibinfo {title} {Financial modelling with jump
  processes}}}\ (\bibinfo  {publisher} {Chapman and Hall/CRC},\ \bibinfo {year}
  {2003})\BibitemShut {NoStop}%
\bibitem [{\citenamefont {Glasserman}(2004)}]{Glasserman2004}%
  \BibitemOpen
  \bibfield  {author} {\bibinfo {author} {\bibfnamefont {P.}~\bibnamefont
  {Glasserman}},\ }\href@noop {} {\emph {\bibinfo {title} {Monte Carlo Methods
  in Financial Engineering}}}\ (\bibinfo  {publisher} {Springer},\ \bibinfo
  {year} {2004})\BibitemShut {NoStop}%
\bibitem [{\citenamefont {Duffie}(2010)}]{duffie2010dynamic}%
  \BibitemOpen
  \bibfield  {author} {\bibinfo {author} {\bibfnamefont {D.}~\bibnamefont
  {Duffie}},\ }\href@noop {} {\emph {\bibinfo {title} {Dynamic asset pricing
  theory}}}\ (\bibinfo  {publisher} {Princeton University Press},\ \bibinfo
  {year} {2010})\BibitemShut {NoStop}%
\bibitem [{\citenamefont {Murphy}(2012)}]{murphy2012machine}%
  \BibitemOpen
  \bibfield  {author} {\bibinfo {author} {\bibfnamefont {K.~P.}\ \bibnamefont
  {Murphy}},\ }\href@noop {} {\emph {\bibinfo {title} {Machine learning: a
  probabilistic perspective}}}\ (\bibinfo  {publisher} {MIT press},\ \bibinfo
  {year} {2012})\BibitemShut {NoStop}%
\bibitem [{\citenamefont {Goodfellow}\ \emph {et~al.}(2016)\citenamefont
  {Goodfellow}, \citenamefont {Bengio}, \citenamefont {Courville},\ and\
  \citenamefont {Bengio}}]{goodfellow2016deep}%
  \BibitemOpen
  \bibfield  {author} {\bibinfo {author} {\bibfnamefont {I.}~\bibnamefont
  {Goodfellow}}, \bibinfo {author} {\bibfnamefont {Y.}~\bibnamefont {Bengio}},
  \bibinfo {author} {\bibfnamefont {A.}~\bibnamefont {Courville}},\ and\
  \bibinfo {author} {\bibfnamefont {Y.}~\bibnamefont {Bengio}},\ }\href@noop {}
  {\emph {\bibinfo {title} {Deep learning}}},\ Vol.~\bibinfo {volume} {1}\
  (\bibinfo  {publisher} {MIT press Cambridge},\ \bibinfo {year}
  {2016})\BibitemShut {NoStop}%
\bibitem [{\citenamefont {Dixon}, \citenamefont {Halperin},\ and\ \citenamefont
  {Bilokon}(2020)}]{Dixon2020}%
  \BibitemOpen
  \bibfield  {author} {\bibinfo {author} {\bibfnamefont {M.~F.}\ \bibnamefont
  {Dixon}}, \bibinfo {author} {\bibfnamefont {I.}~\bibnamefont {Halperin}},\
  and\ \bibinfo {author} {\bibfnamefont {P.}~\bibnamefont {Bilokon}},\
  }\href@noop {} {\emph {\bibinfo {title} {Machine Learning in Finance: From
  Theory to Practice}}}\ (\bibinfo  {publisher} {Springer},\ \bibinfo {year}
  {2020})\BibitemShut {NoStop}%
\bibitem [{\citenamefont {Liu}, \citenamefont {Oosterlee},\ and\ \citenamefont
  {Bohte}(2019)}]{Liu2019}%
  \BibitemOpen
  \bibfield  {author} {\bibinfo {author} {\bibfnamefont {S.}~\bibnamefont
  {Liu}}, \bibinfo {author} {\bibfnamefont {C.~W.}\ \bibnamefont {Oosterlee}},\
  and\ \bibinfo {author} {\bibfnamefont {S.~M.}\ \bibnamefont {Bohte}},\
  }\bibfield  {title} {\enquote {\bibinfo {title} {Pricing options and
  computing implied volatilities using neural networks},}\ }\href@noop {}
  {\bibfield  {journal} {\bibinfo  {journal} {Risks}\ }\textbf {\bibinfo
  {volume} {7}},\ \bibinfo {pages} {16} (\bibinfo {year} {2019})}\BibitemShut
  {NoStop}%
\bibitem [{\citenamefont {Hirsa}, \citenamefont {Karatas},\ and\ \citenamefont
  {Oskoui}(2019)}]{Hirsa2019}%
  \BibitemOpen
  \bibfield  {author} {\bibinfo {author} {\bibfnamefont {A.}~\bibnamefont
  {Hirsa}}, \bibinfo {author} {\bibfnamefont {T.}~\bibnamefont {Karatas}},\
  and\ \bibinfo {author} {\bibfnamefont {A.}~\bibnamefont {Oskoui}},\
  }\bibfield  {title} {\enquote {\bibinfo {title} {Supervised deep neural
  networks (dnns) for pricing/calibration of vanilla/exotic options},}\
  }\href@noop {} {\bibfield  {journal} {\bibinfo  {journal} {SSRN Electronic
  Journal}\ } (\bibinfo {year} {2019})}\BibitemShut {NoStop}%
\bibitem [{\citenamefont {Raissi}, \citenamefont {Perdikaris},\ and\
  \citenamefont {Karniadakis}(2019)}]{raissi2019physics}%
  \BibitemOpen
  \bibfield  {author} {\bibinfo {author} {\bibfnamefont {M.}~\bibnamefont
  {Raissi}}, \bibinfo {author} {\bibfnamefont {P.}~\bibnamefont {Perdikaris}},\
  and\ \bibinfo {author} {\bibfnamefont {G.~E.}\ \bibnamefont {Karniadakis}},\
  }\bibfield  {title} {\enquote {\bibinfo {title} {Physics-informed neural
  networks: A deep learning framework for solving forward and inverse problems
  involving nonlinear partial differential equations},}\ }\href@noop {}
  {\bibfield  {journal} {\bibinfo  {journal} {Journal of Computational
  physics}\ }\textbf {\bibinfo {volume} {378}},\ \bibinfo {pages} {686--707}
  (\bibinfo {year} {2019})}\BibitemShut {NoStop}%
\bibitem [{\citenamefont {Gatta}\ \emph {et~al.}(2023)\citenamefont {Gatta},
  \citenamefont {Di~Cola}, \citenamefont {Giampaolo}, \citenamefont
  {Piccialli},\ and\ \citenamefont {Cuomo}}]{gatta2023meshless}%
  \BibitemOpen
  \bibfield  {author} {\bibinfo {author} {\bibfnamefont {F.}~\bibnamefont
  {Gatta}}, \bibinfo {author} {\bibfnamefont {V.~S.}\ \bibnamefont {Di~Cola}},
  \bibinfo {author} {\bibfnamefont {F.}~\bibnamefont {Giampaolo}}, \bibinfo
  {author} {\bibfnamefont {F.}~\bibnamefont {Piccialli}},\ and\ \bibinfo
  {author} {\bibfnamefont {S.}~\bibnamefont {Cuomo}},\ }\bibfield  {title}
  {\enquote {\bibinfo {title} {Meshless methods for american option pricing
  through physics-informed neural networks},}\ }\href@noop {} {\bibfield
  {journal} {\bibinfo  {journal} {Engineering Analysis with Boundary Elements}\
  }\textbf {\bibinfo {volume} {151}},\ \bibinfo {pages} {68--82} (\bibinfo
  {year} {2023})}\BibitemShut {NoStop}%
\bibitem [{\citenamefont {Hainaut}\ and\ \citenamefont
  {Casas}(2024)}]{hainaut2024option}%
  \BibitemOpen
  \bibfield  {author} {\bibinfo {author} {\bibfnamefont {D.}~\bibnamefont
  {Hainaut}}\ and\ \bibinfo {author} {\bibfnamefont {A.}~\bibnamefont
  {Casas}},\ }\bibfield  {title} {\enquote {\bibinfo {title} {Option pricing in
  the heston model with physics inspired neural networks},}\ }\href@noop {}
  {\bibfield  {journal} {\bibinfo  {journal} {Annals of Finance}\ }\textbf
  {\bibinfo {volume} {20}},\ \bibinfo {pages} {353--376} (\bibinfo {year}
  {2024})}\BibitemShut {NoStop}%
\bibitem [{\citenamefont {Wang}, \citenamefont {Li},\ and\ \citenamefont
  {Li}(2023)}]{wang2023deep}%
  \BibitemOpen
  \bibfield  {author} {\bibinfo {author} {\bibfnamefont {X.}~\bibnamefont
  {Wang}}, \bibinfo {author} {\bibfnamefont {J.}~\bibnamefont {Li}},\ and\
  \bibinfo {author} {\bibfnamefont {J.}~\bibnamefont {Li}},\ }\bibfield
  {title} {\enquote {\bibinfo {title} {A deep learning based numerical pde
  method for option pricing},}\ }\href@noop {} {\bibfield  {journal} {\bibinfo
  {journal} {Computational economics}\ }\textbf {\bibinfo {volume} {62}},\
  \bibinfo {pages} {149--164} (\bibinfo {year} {2023})}\BibitemShut {NoStop}%
\bibitem [{\citenamefont {Bai}, \citenamefont {Chaolu},\ and\ \citenamefont
  {Bilige}(2022)}]{bai2022application}%
  \BibitemOpen
  \bibfield  {author} {\bibinfo {author} {\bibfnamefont {Y.}~\bibnamefont
  {Bai}}, \bibinfo {author} {\bibfnamefont {T.}~\bibnamefont {Chaolu}},\ and\
  \bibinfo {author} {\bibfnamefont {S.}~\bibnamefont {Bilige}},\ }\bibfield
  {title} {\enquote {\bibinfo {title} {The application of improved
  physics-informed neural network (ipinn) method in finance},}\ }\href@noop {}
  {\bibfield  {journal} {\bibinfo  {journal} {Nonlinear Dynamics}\ }\textbf
  {\bibinfo {volume} {107}},\ \bibinfo {pages} {3655--3667} (\bibinfo {year}
  {2022})}\BibitemShut {NoStop}%
\bibitem [{\citenamefont {De~Spiegeleer}\ \emph {et~al.}(2018)\citenamefont
  {De~Spiegeleer}, \citenamefont {Madan}, \citenamefont {Reyners},\ and\
  \citenamefont {Schoutens}}]{de2018machine}%
  \BibitemOpen
  \bibfield  {author} {\bibinfo {author} {\bibfnamefont {J.}~\bibnamefont
  {De~Spiegeleer}}, \bibinfo {author} {\bibfnamefont {D.~B.}\ \bibnamefont
  {Madan}}, \bibinfo {author} {\bibfnamefont {S.}~\bibnamefont {Reyners}},\
  and\ \bibinfo {author} {\bibfnamefont {W.}~\bibnamefont {Schoutens}},\
  }\bibfield  {title} {\enquote {\bibinfo {title} {Machine learning for
  quantitative finance: fast derivative pricing, hedging and fitting},}\
  }\href@noop {} {\bibfield  {journal} {\bibinfo  {journal} {Quantitative
  Finance}\ }\textbf {\bibinfo {volume} {18}},\ \bibinfo {pages} {1635--1643}
  (\bibinfo {year} {2018})}\BibitemShut {NoStop}%
\bibitem [{\citenamefont {Ndikum}(2020)}]{ndikum2020machine}%
  \BibitemOpen
  \bibfield  {author} {\bibinfo {author} {\bibfnamefont {P.}~\bibnamefont
  {Ndikum}},\ }\bibfield  {title} {\enquote {\bibinfo {title} {Machine learning
  algorithms for financial asset price forecasting},}\ }\href@noop {}
  {\bibfield  {journal} {\bibinfo  {journal} {arXiv preprint arXiv:2004.01504}\
  } (\bibinfo {year} {2020})}\BibitemShut {NoStop}%
\bibitem [{\citenamefont {Gaspar}, \citenamefont {Lopes},\ and\ \citenamefont
  {Sequeira}(2020)}]{gaspar2020neural}%
  \BibitemOpen
  \bibfield  {author} {\bibinfo {author} {\bibfnamefont {R.~M.}\ \bibnamefont
  {Gaspar}}, \bibinfo {author} {\bibfnamefont {S.~D.}\ \bibnamefont {Lopes}},\
  and\ \bibinfo {author} {\bibfnamefont {B.}~\bibnamefont {Sequeira}},\
  }\bibfield  {title} {\enquote {\bibinfo {title} {Neural network pricing of
  american put options},}\ }\href@noop {} {\bibfield  {journal} {\bibinfo
  {journal} {Risks}\ }\textbf {\bibinfo {volume} {8}},\ \bibinfo {pages} {73}
  (\bibinfo {year} {2020})}\BibitemShut {NoStop}%
\bibitem [{\citenamefont {Anderson}\ and\ \citenamefont
  {Ulrych}(2023)}]{anderson2023accelerated}%
  \BibitemOpen
  \bibfield  {author} {\bibinfo {author} {\bibfnamefont {D.}~\bibnamefont
  {Anderson}}\ and\ \bibinfo {author} {\bibfnamefont {U.}~\bibnamefont
  {Ulrych}},\ }\bibfield  {title} {\enquote {\bibinfo {title} {Accelerated
  american option pricing with deep neural networks},}\ }\href@noop {}
  {\bibfield  {journal} {\bibinfo  {journal} {Quantitative Finance and
  Economics}\ }\textbf {\bibinfo {volume} {7}},\ \bibinfo {pages} {207--228}
  (\bibinfo {year} {2023})}\BibitemShut {NoStop}%
\bibitem [{\citenamefont {Cao}\ \emph {et~al.}(2021)\citenamefont {Cao},
  \citenamefont {Chen}, \citenamefont {Hull},\ and\ \citenamefont
  {Poulos}}]{cao2021deep}%
  \BibitemOpen
  \bibfield  {author} {\bibinfo {author} {\bibfnamefont {J.}~\bibnamefont
  {Cao}}, \bibinfo {author} {\bibfnamefont {J.}~\bibnamefont {Chen}}, \bibinfo
  {author} {\bibfnamefont {J.}~\bibnamefont {Hull}},\ and\ \bibinfo {author}
  {\bibfnamefont {Z.}~\bibnamefont {Poulos}},\ }\bibfield  {title} {\enquote
  {\bibinfo {title} {Deep learning for exotic option valuation},}\ }\href@noop
  {} {\bibfield  {journal} {\bibinfo  {journal} {arXiv preprint
  arXiv:2103.12551}\ } (\bibinfo {year} {2021})}\BibitemShut {NoStop}%
\bibitem [{\citenamefont {Ruf}\ and\ \citenamefont
  {Wang}(2019)}]{ruf2019neural}%
  \BibitemOpen
  \bibfield  {author} {\bibinfo {author} {\bibfnamefont {J.}~\bibnamefont
  {Ruf}}\ and\ \bibinfo {author} {\bibfnamefont {W.}~\bibnamefont {Wang}},\
  }\bibfield  {title} {\enquote {\bibinfo {title} {Neural networks for option
  pricing and hedging: a literature review},}\ }\href@noop {} {\bibfield
  {journal} {\bibinfo  {journal} {arXiv preprint arXiv:1911.05620}\ } (\bibinfo
  {year} {2019})}\BibitemShut {NoStop}%
\bibitem [{\citenamefont {Bloch}(2019)}]{bloch2019option}%
  \BibitemOpen
  \bibfield  {author} {\bibinfo {author} {\bibfnamefont {D.~A.}\ \bibnamefont
  {Bloch}},\ }\bibfield  {title} {\enquote {\bibinfo {title} {Option pricing
  with machine learning},}\ }\href@noop {} {\bibfield  {journal} {\bibinfo
  {journal} {Available at SSRN 3486224}\ } (\bibinfo {year}
  {2019})}\BibitemShut {NoStop}%
\bibitem [{\citenamefont {Culkin}\ and\ \citenamefont
  {Das}(2017)}]{culkin2017machine}%
  \BibitemOpen
  \bibfield  {author} {\bibinfo {author} {\bibfnamefont {R.}~\bibnamefont
  {Culkin}}\ and\ \bibinfo {author} {\bibfnamefont {S.~R.}\ \bibnamefont
  {Das}},\ }\bibfield  {title} {\enquote {\bibinfo {title} {Machine learning in
  finance: the case of deep learning for option pricing},}\ }\href@noop {}
  {\bibfield  {journal} {\bibinfo  {journal} {Journal of Investment
  Management}\ }\textbf {\bibinfo {volume} {15}},\ \bibinfo {pages} {92--100}
  (\bibinfo {year} {2017})}\BibitemShut {NoStop}%
\bibitem [{\citenamefont {Doersch}(2016)}]{doersch2016tutorial}%
  \BibitemOpen
  \bibfield  {author} {\bibinfo {author} {\bibfnamefont {C.}~\bibnamefont
  {Doersch}},\ }\bibfield  {title} {\enquote {\bibinfo {title} {Tutorial on
  variational autoencoders},}\ }\href@noop {} {\bibfield  {journal} {\bibinfo
  {journal} {arXiv preprint arXiv:1606.05908}\ } (\bibinfo {year}
  {2016})}\BibitemShut {NoStop}%
\bibitem [{\citenamefont {Pu}\ \emph {et~al.}(2016)\citenamefont {Pu},
  \citenamefont {Gan}, \citenamefont {Henao}, \citenamefont {Yuan},
  \citenamefont {Li}, \citenamefont {Stevens},\ and\ \citenamefont
  {Carin}}]{pu2016variational}%
  \BibitemOpen
  \bibfield  {author} {\bibinfo {author} {\bibfnamefont {Y.}~\bibnamefont
  {Pu}}, \bibinfo {author} {\bibfnamefont {Z.}~\bibnamefont {Gan}}, \bibinfo
  {author} {\bibfnamefont {R.}~\bibnamefont {Henao}}, \bibinfo {author}
  {\bibfnamefont {X.}~\bibnamefont {Yuan}}, \bibinfo {author} {\bibfnamefont
  {C.}~\bibnamefont {Li}}, \bibinfo {author} {\bibfnamefont {A.}~\bibnamefont
  {Stevens}},\ and\ \bibinfo {author} {\bibfnamefont {L.}~\bibnamefont
  {Carin}},\ }\bibfield  {title} {\enquote {\bibinfo {title} {Variational
  autoencoder for deep learning of images, labels and captions},}\ }\href@noop
  {} {\bibfield  {journal} {\bibinfo  {journal} {Advances in neural information
  processing systems}\ }\textbf {\bibinfo {volume} {29}} (\bibinfo {year}
  {2016})}\BibitemShut {NoStop}%
\bibitem [{\citenamefont {Bergeron}\ \emph {et~al.}(2021)\citenamefont
  {Bergeron}, \citenamefont {Fung}, \citenamefont {Hull},\ and\ \citenamefont
  {Poulos}}]{bergeron2021variational}%
  \BibitemOpen
  \bibfield  {author} {\bibinfo {author} {\bibfnamefont {M.}~\bibnamefont
  {Bergeron}}, \bibinfo {author} {\bibfnamefont {N.}~\bibnamefont {Fung}},
  \bibinfo {author} {\bibfnamefont {J.}~\bibnamefont {Hull}},\ and\ \bibinfo
  {author} {\bibfnamefont {Z.}~\bibnamefont {Poulos}},\ }\bibfield  {title}
  {\enquote {\bibinfo {title} {Variational autoencoders: A hands-off approach
  to volatility},}\ }\href@noop {} {\bibfield  {journal} {\bibinfo  {journal}
  {arXiv preprint arXiv:2102.03945}\ } (\bibinfo {year} {2021})}\BibitemShut
  {NoStop}%
\bibitem [{\citenamefont {Vecer}(2001)}]{vecer2001new}%
  \BibitemOpen
  \bibfield  {author} {\bibinfo {author} {\bibfnamefont {J.}~\bibnamefont
  {Vecer}},\ }\bibfield  {title} {\enquote {\bibinfo {title} {A new pde
  approach for pricing arithmetic average asian options},}\ }\href@noop {}
  {\bibfield  {journal} {\bibinfo  {journal} {Journal of computational
  finance}\ }\textbf {\bibinfo {volume} {4}},\ \bibinfo {pages} {105--113}
  (\bibinfo {year} {2001})}\BibitemShut {NoStop}%
\bibitem [{\citenamefont {Gatheral}\ and\ \citenamefont
  {Jacquier}(2014)}]{gatheral2014arbitrage}%
  \BibitemOpen
  \bibfield  {author} {\bibinfo {author} {\bibfnamefont {J.}~\bibnamefont
  {Gatheral}}\ and\ \bibinfo {author} {\bibfnamefont {A.}~\bibnamefont
  {Jacquier}},\ }\bibfield  {title} {\enquote {\bibinfo {title} {Arbitrage-free
  svi volatility surfaces},}\ }\href@noop {} {\bibfield  {journal} {\bibinfo
  {journal} {Quantitative Finance}\ }\textbf {\bibinfo {volume} {14}},\
  \bibinfo {pages} {59--71} (\bibinfo {year} {2014})}\BibitemShut {NoStop}%
\bibitem [{\citenamefont {Ding}, \citenamefont {Lu},\ and\ \citenamefont
  {Cheung}(2025)}]{ding2025fast}%
  \BibitemOpen
  \bibfield  {author} {\bibinfo {author} {\bibfnamefont {L.}~\bibnamefont
  {Ding}}, \bibinfo {author} {\bibfnamefont {E.}~\bibnamefont {Lu}},\ and\
  \bibinfo {author} {\bibfnamefont {K.}~\bibnamefont {Cheung}},\ }\bibfield
  {title} {\enquote {\bibinfo {title} {Fast derivative valuation from
  volatility surfaces using machine learning},}\ }\href@noop {} {\bibfield
  {journal} {\bibinfo  {journal} {arXiv preprint arXiv:2505.22957}\ } (\bibinfo
  {year} {2025})}\BibitemShut {NoStop}%
\bibitem [{\citenamefont {Varma}\ and\ \citenamefont
  {Virmani}(2015)}]{varma2015derivatives}%
  \BibitemOpen
  \bibfield  {author} {\bibinfo {author} {\bibfnamefont {J.~R.}\ \bibnamefont
  {Varma}}\ and\ \bibinfo {author} {\bibfnamefont {V.}~\bibnamefont
  {Virmani}},\ }\href@noop {} {\emph {\bibinfo {title} {Derivatives Pricing
  using QuantLib: An Introduction}}}\ (\bibinfo  {publisher} {Indian Institute
  of Management},\ \bibinfo {year} {2015})\BibitemShut {NoStop}%
\bibitem [{\citenamefont {Varma}\ and\ \citenamefont
  {Virmani}(2016)}]{varma2016computational}%
  \BibitemOpen
  \bibfield  {author} {\bibinfo {author} {\bibfnamefont {J.~R.}\ \bibnamefont
  {Varma}}\ and\ \bibinfo {author} {\bibfnamefont {V.}~\bibnamefont
  {Virmani}},\ }\bibfield  {title} {\enquote {\bibinfo {title} {Computational
  finance using quantlib-python},}\ }\href@noop {} {\bibfield  {journal}
  {\bibinfo  {journal} {Computing in Science \& Engineering}\ }\textbf
  {\bibinfo {volume} {18}},\ \bibinfo {pages} {78--88} (\bibinfo {year}
  {2016})}\BibitemShut {NoStop}%
\bibitem [{\citenamefont {Ackerer}, \citenamefont {Tagasovska},\ and\
  \citenamefont {Vatter}(2020)}]{ackerer2020deep}%
  \BibitemOpen
  \bibfield  {author} {\bibinfo {author} {\bibfnamefont {D.}~\bibnamefont
  {Ackerer}}, \bibinfo {author} {\bibfnamefont {N.}~\bibnamefont
  {Tagasovska}},\ and\ \bibinfo {author} {\bibfnamefont {T.}~\bibnamefont
  {Vatter}},\ }\bibfield  {title} {\enquote {\bibinfo {title} {Deep smoothing
  of the implied volatility surface},}\ }\href@noop {} {\bibfield  {journal}
  {\bibinfo  {journal} {Advances in Neural Information Processing Systems}\
  }\textbf {\bibinfo {volume} {33}},\ \bibinfo {pages} {11552--11563} (\bibinfo
  {year} {2020})}\BibitemShut {NoStop}%
\bibitem [{\citenamefont {Ning}\ \emph {et~al.}(2023)\citenamefont {Ning},
  \citenamefont {Jaimungal}, \citenamefont {Zhang},\ and\ \citenamefont
  {Bergeron}}]{ning2023arbitrage}%
  \BibitemOpen
  \bibfield  {author} {\bibinfo {author} {\bibfnamefont {B.}~\bibnamefont
  {Ning}}, \bibinfo {author} {\bibfnamefont {S.}~\bibnamefont {Jaimungal}},
  \bibinfo {author} {\bibfnamefont {X.}~\bibnamefont {Zhang}},\ and\ \bibinfo
  {author} {\bibfnamefont {M.}~\bibnamefont {Bergeron}},\ }\bibfield  {title}
  {\enquote {\bibinfo {title} {Arbitrage-free implied volatility surface
  generation with variational autoencoders},}\ }\href@noop {} {\bibfield
  {journal} {\bibinfo  {journal} {SIAM Journal on Financial Mathematics}\
  }\textbf {\bibinfo {volume} {14}},\ \bibinfo {pages} {1004--1027} (\bibinfo
  {year} {2023})}\BibitemShut {NoStop}%
\bibitem [{\citenamefont {Wang}, \citenamefont {Liu},\ and\ \citenamefont
  {Vuik}(2025)}]{wang2025controllable}%
  \BibitemOpen
  \bibfield  {author} {\bibinfo {author} {\bibfnamefont {J.}~\bibnamefont
  {Wang}}, \bibinfo {author} {\bibfnamefont {S.}~\bibnamefont {Liu}},\ and\
  \bibinfo {author} {\bibfnamefont {C.}~\bibnamefont {Vuik}},\ }\bibfield
  {title} {\enquote {\bibinfo {title} {Controllable generation of implied
  volatility surfaces with variational autoencoders},}\ }\href@noop {}
  {\bibfield  {journal} {\bibinfo  {journal} {arXiv preprint arXiv:2509.01743}\
  } (\bibinfo {year} {2025})}\BibitemShut {NoStop}%
\bibitem [{\citenamefont {Kelly}\ \emph {et~al.}(2023)\citenamefont {Kelly},
  \citenamefont {Kuznetsov}, \citenamefont {Malamud},\ and\ \citenamefont
  {Xu}}]{kelly2023deep}%
  \BibitemOpen
  \bibfield  {author} {\bibinfo {author} {\bibfnamefont {B.~T.}\ \bibnamefont
  {Kelly}}, \bibinfo {author} {\bibfnamefont {B.}~\bibnamefont {Kuznetsov}},
  \bibinfo {author} {\bibfnamefont {S.}~\bibnamefont {Malamud}},\ and\ \bibinfo
  {author} {\bibfnamefont {T.~A.}\ \bibnamefont {Xu}},\ }\bibfield  {title}
  {\enquote {\bibinfo {title} {Deep learning from implied volatility
  surfaces},}\ }\href@noop {} {\bibfield  {journal} {\bibinfo  {journal} {Swiss
  Finance Institute Research Paper}\ } (\bibinfo {year} {2023})}\BibitemShut
  {NoStop}%
\bibitem [{\citenamefont {Fan}, \citenamefont {Gupta},\ and\ \citenamefont
  {Ritchken}(2003)}]{fan2003hedging}%
  \BibitemOpen
  \bibfield  {author} {\bibinfo {author} {\bibfnamefont {R.}~\bibnamefont
  {Fan}}, \bibinfo {author} {\bibfnamefont {A.}~\bibnamefont {Gupta}},\ and\
  \bibinfo {author} {\bibfnamefont {P.}~\bibnamefont {Ritchken}},\ }\bibfield
  {title} {\enquote {\bibinfo {title} {Hedging in the possible presence of
  unspanned stochastic volatility: Evidence from swaption markets},}\
  }\href@noop {} {\bibfield  {journal} {\bibinfo  {journal} {The Journal of
  Finance}\ }\textbf {\bibinfo {volume} {58}},\ \bibinfo {pages} {2219--2248}
  (\bibinfo {year} {2003})}\BibitemShut {NoStop}%
\bibitem [{\citenamefont {Hagan}, \citenamefont {Lesniewski},\ and\
  \citenamefont {Woodward}(2015)}]{hagan2015probability}%
  \BibitemOpen
  \bibfield  {author} {\bibinfo {author} {\bibfnamefont {P.}~\bibnamefont
  {Hagan}}, \bibinfo {author} {\bibfnamefont {A.}~\bibnamefont {Lesniewski}},\
  and\ \bibinfo {author} {\bibfnamefont {D.}~\bibnamefont {Woodward}},\
  }\bibfield  {title} {\enquote {\bibinfo {title} {Probability distribution in
  the sabr model of stochastic volatility},}\ }in\ \href@noop {} {\emph
  {\bibinfo {booktitle} {Large deviations and asymptotic methods in finance}}}\
  (\bibinfo  {publisher} {Springer},\ \bibinfo {year} {2015})\ pp.\ \bibinfo
  {pages} {1--35}\BibitemShut {NoStop}%
\end{thebibliography}%

\onecolumngrid
\clearpage
\appendix

\end{document}